\documentclass[10pt,preprint]{aastex}

\def\Ell{{\cal L}}
\def\bi{\begin{itemize}}
\def\ei{\end{itemize}}
\def\be{\begin{equation}}
\def\ee{\end{equation}}

\def\gtrsim{\mathrel{\hbox{\rlap{\hbox{\lower4pt\hbox{$\sim$}}}\hbox{$>$}}}}
\def\lesssim{\mathrel{\hbox{\rlap{\hbox{\lower4pt\hbox{$\sim$}}}\hbox{$<$}}}}
\def\gtrsim{\mathrel{\hbox{\rlap{\hbox{\lower4pt\hbox{$\sim$}}}\hbox{$>$}}}}
\def\lesssim{\mathrel{\hbox{\rlap{\hbox{\lower4pt\hbox{$\sim$}}}\hbox{$<$}}}}

\slugcomment{Accepted for publication in the {\it Astrophysical Journal}.}
\shortauthors{Lister}
\shorttitle{Altered Luminosity Functions of Relativistic Jets}
\received{2003 May 13}
\accepted{2003 Aug 29}
\begin{document}
\title{Altered Luminosity Functions of Relativistically
Beamed Jet Populations}

\author{M. L. Lister\footnote{Karl Jansky Postdoctoral Fellow}}

\affil{National Radio Astronomy Observatory, 520 Edgemont
Road, Charlottesville, VA 22903--2454 \footnote{Present
Address: Department of Physics, Purdue University, 525 Northwestern
Avenue, West Lafayette, IN 47907}}

\email{mlister@nrao.edu}

\begin{abstract}
The intrinsic luminosity functions of extremely fast jets found in
many active galaxies and gamma-ray bursts are difficult to measure
since their apparent luminosities are strongly affected by
relativistic beaming. Past studies have only provided analytical predictions
for the beamed characteristics of populations in which all jets
have the same Lorentz factor. However, the jets found in active
galaxies are known to span a large range of speeds.  Here we derive
analytical expressions for the expected Doppler factor distributions
and apparent (beamed) luminosity functions of randomly-oriented,
two-sided jet populations which have bulk Lorentz factors distributed
according to a simple power-law over a range [G1, G2].
We find that if a jet population has a uniform or inverted Lorentz
factor distribution, its Doppler factor distribution and beamed
luminosity function will be very similar to that of a jet population
with Lorentz factors all equal to G2. In particular, the
beamed and intrinsic luminosity functions will have the same slope at
high luminosities.  In the case of a steeply declining Lorentz factor
distribution, the slopes at high luminosities will also be identical,
but only up to an apparent luminosity equivalent to the upper
cutoff of the intrinsic luminosity function. Also, at very high apparent
luminosities, the slope will be proportional to the power law slope of
the Lorentz factor distribution. At very low luminosities, the form of
the apparent luminosity function is very sensitive to the lower cutoff
and steepness of the Lorentz factor distribution. We discuss how it
should be possible to recover useful information about the intrinsic
luminosity functions of relativistic jets by studying flux-limited
samples selected on the basis of beamed jet emission.

\end{abstract}


\section{INTRODUCTION}

The observed fluxes of highly relativistic jets such as those found in
active galactic nuclei and gamma-ray bursts are subject to
strong relativistic aberration and Doppler beaming effects. This poses
a serious challenge for determining the intrinsic
luminosity functions (LFs) of these objects. Previous studies (e.g.,
\citealt*{JW99, US84, UP91, UPS91, UP95}) have discussed in detail how
luminosity functions can be apparently altered by Doppler beaming, but
present analytical expressions only for cases where every jet in
the population has the same intrinsic jet speed. In their analysis of
beamed luminosity functions, \cite*{UPS91} found evidence that BL Lac objects could
be the beamed counterparts of FR-I galaxies, provided that
they have a wide distribution of jet Lorentz factors.  Large proper
motion surveys \citep{V03,K03} have also
revealed that the overall AGN population contains a wide range of apparent
jet speeds, which is inconsistent with a single-valued intrinsic
speed distribution
\citep{VC94,LM97}.

Here we extend the earlier work of \cite{UP91} by deriving full
analytical expressions for the Doppler factor distributions and beamed
luminosity functions of randomly-oriented, two-sided jet populations
that contain a range of jet speed $v$. We use an improved method that
involves first deriving the predicted Doppler factor distribution, and
then integrating over an appropriate range of Doppler factor to obtain
the apparent luminosity function. We show how this method
results in a much wider number of scenarios that can be treated fully
analytically. 

For simplicity, we consider only straight jets, where
the emission is beamed proportionally according to the Lorentz factor
of the jet flow (defined as $\gamma = (1-(v/c)^2)^{-1/2}$, where $c$
is the speed of light). We restrict our analysis to cases where the
jet Lorentz factors are distributed according to a power law, since
this distribution provides the best fit to the apparent speed
distribution of AGN jets
\citep{LM97}. Other forms, such as Gaussian and single-valued
functions, provide very poor fits to the observed speeds. 

In \S~\ref{dopplersec} we outline our model assumptions and derive the
predicted Doppler factor distributions for Lorentz factor
distributions of the form $P_{\gamma} \propto \gamma^a$. We use these results
in \S~\ref{beamedlfsec} to examine how a simple intrinsic power-law
luminosity function $\phi(\Ell) \propto \Ell^{-B}$ is transformed by
Doppler boosting into its observed form $\Phi(L)$. In
\S~\ref{obsconseq} we discuss how our results can be applied to
recover useful information about the parent population of samples
dominated by relativistic jet emission, such as gamma-ray bursts,
gamma-ray loud AGN, and core-dominated, radio-loud AGN. We summarize our
findings in \S~\ref{summary}.

\section{DOPPLER BEAMING FACTOR DISTRIBUTIONS\label{dopplersec}}

For the purposes of obtaining analytical expressions, we adopt a 
simple model in which each source in the population contains two
identical, straight, oppositely-directed jets with bulk Lorentz factor
$\gamma$.  We assume that the jet axes are randomly oriented, so that if $\theta$ represents the angle between
the  jet axis and the line of sight, then $P_{\theta} =
\sin{\theta}$, and  $0\arcdeg \le \theta \le 90\arcdeg$, where $P_{\theta}$
represents the probability density function of the viewing angles.
The Lorentz factors have a density function $P_{\gamma}$ that is
positive and non-zero over the region $\gamma_1 \le \gamma \le
\gamma_2$.  The kinematic Doppler factor is defined as \be \delta =
\left[\gamma - \sqrt{\gamma^2-1}\cos{\theta}\right]^{-1} ,\ee and
represents the frequency shift between the observer and jet rest
frames.  The minimum possible Doppler factor is $\delta_{min} =
\gamma_2^{-1}$, and occurs when $\theta = 90\arcdeg$.  The maximum
possible Doppler factor in the population is $\delta_{max} = \gamma_2
(1 + \beta_2)$, where $\beta = v/c$, and occurs when $\theta =
0\arcdeg$. When $\gamma_2$ is large, $\delta_{max} \simeq 2\gamma_2$.

In Appendix A we derive the expected Doppler factor distributions of
randomly-oriented jet populations having various Lorentz factor
distributions. We plot some individual cases corresponding to
single-valued and power-law distributions in Figure~\ref{fi1}. 

\subsection{Single-Valued Lorentz Factor Distributions}

 The two thick lines in Figure~\ref{fi1} represent populations where
 all jets have the same intrinsic Lorentz factor $\gamma_o$, with the
 dotted-dashed line corresponding to $\gamma_o = 3$, and the solid
 line corresponding to $\gamma_o = 30$. The latter represents the
 fastest jet speed typically seen in large AGN proper-motion surveys
 (Kellermann et al., in preparation). For a single-valued Lorentz
 factor distribution, $P_{\delta} \propto
\delta^{-2}$, which corresponds to a straight line with slope of $-2$
in our log-log plot (Fig.~\ref{fi1}).

\subsection{Power-Law Lorentz Factor Distributions}
\begin{figure*}
\epsscale{0.9}
\plotone{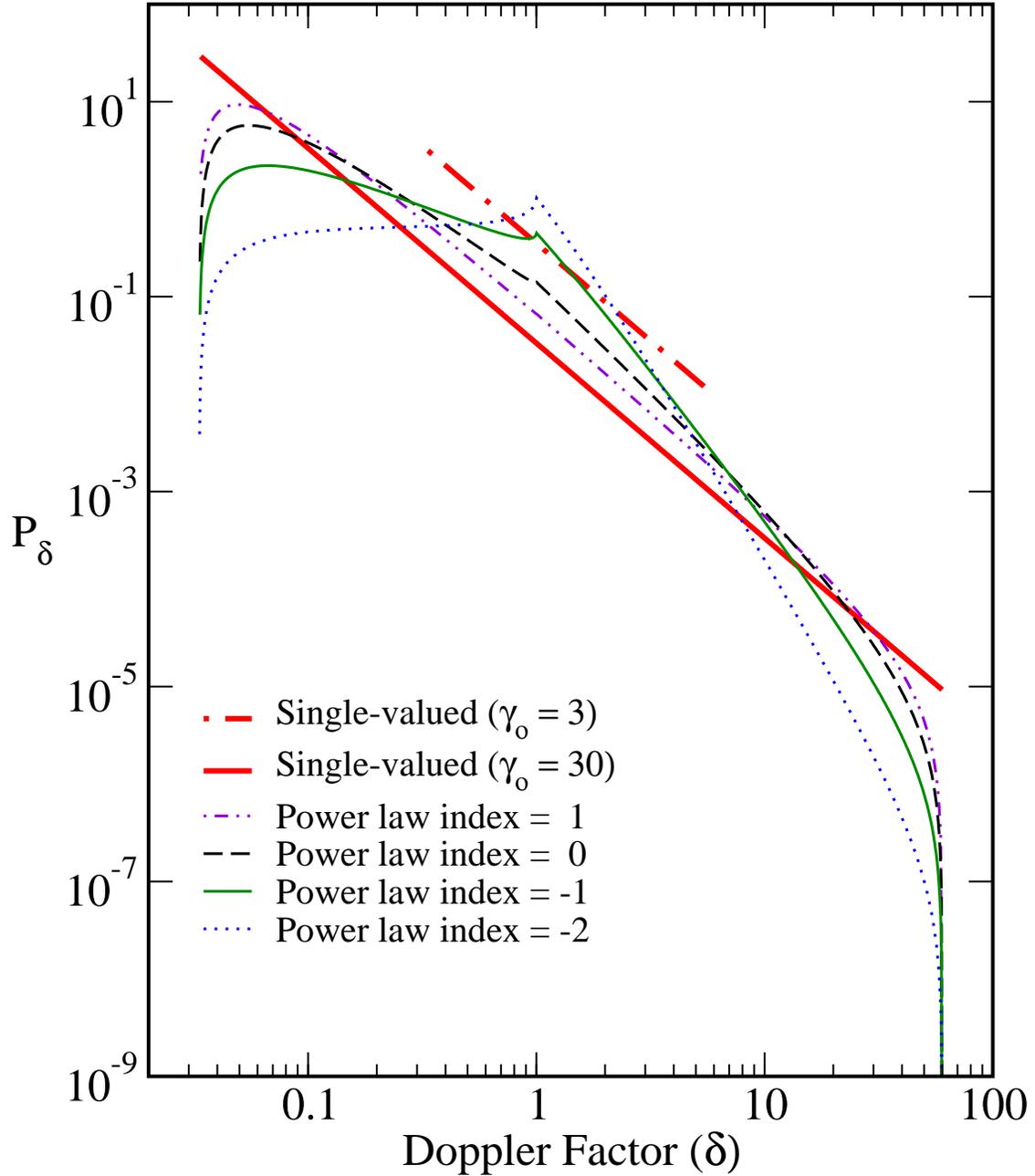}
\caption{\label{fi1} Predicted Doppler boosting factor distribution
functions for randomly-oriented populations of two-sided jets. The
thick dot-dashed and dotted lines correspond to a jet population having
a single bulk Lorentz factor of  $\gamma_o = 3$ and $\gamma_o = 30$,
respectively. The thin lines correspond to jet populations whose
bulk  Lorentz factors are distributed over the range $1 \le \gamma \le
30$ with a density function $P_{\gamma} \propto \gamma^{a}$. The
models plotted correspond to $a = 1$ (thin dot-dashed line), $a = 0$ (thin
dashed line), $a = -1$ (thin solid line), and $a = -2$ (thin dotted line).  }
\end{figure*}

The thin curves in Figure~\ref{fi1} correspond to various jet populations
with Lorentz factors distributed between $1 \le \gamma \le
30$. The thin dotted-dashed line shows the expected Doppler factor distribution for a
jet population with an inverted $\gamma$ distribution (i.e.,
$P_{\gamma} \propto \gamma$). The thin dashed curve represents a uniform
$\gamma$ distribution ($P_{\gamma} =$ constant). These two curves have
an overall slope that is roughly the same as in the single-valued
case (slope = $-2$) over a wide range of $\delta$, but they both
taper off sharply near $\delta_{min}$ and $\delta_{max}$.  Jets with
these extreme Doppler factors all have Lorentz factors $\simeq
\gamma_2$, and viewing angles close to $90\arcdeg$ and $0\arcdeg$,
respectively. They are therefore very rare in the population. In the
case of models where $P_{\gamma}
\propto \gamma^a$ and $a \le 0$, the tapering off at the extrema is a result of 
having relatively fewer jets with $\gamma = \gamma_2$, as compared to a
single-valued population where all the jets have the same jet speed.

The remaining two curves in Figure~\ref{fi1} represent the cases
$P_{\gamma} \propto \gamma^{-1}$ (thin solid line) and $P_{\gamma}
\propto \gamma^{-2}$ (thin dotted line).  In populations such as
these which have few high-speed jets, the distributions taper
off much more sharply for Doppler factors above $\gamma_1^{-1}$. 

The slopes of the $P_{\delta}$ functions have a
discontinuity at $\delta = \gamma_1^{-1}$, which arises because, if
$\gamma_2 > \gamma_1(1+\beta_1)$, potentially any jet in the population can have a
Doppler factor in the range $\gamma_1^{-1} <
\delta < \gamma_1(1+\beta_1)$. This leads to a slight over-density of
sources in this interval. For a jet to have a Doppler factor that lies
outside this range, it must have a speed greater than (but not equal
to) $\beta_1$.

\subsection{Discussion\label{deltadiscussion}}

\begin{figure*}
\epsscale{0.9}
\plotone{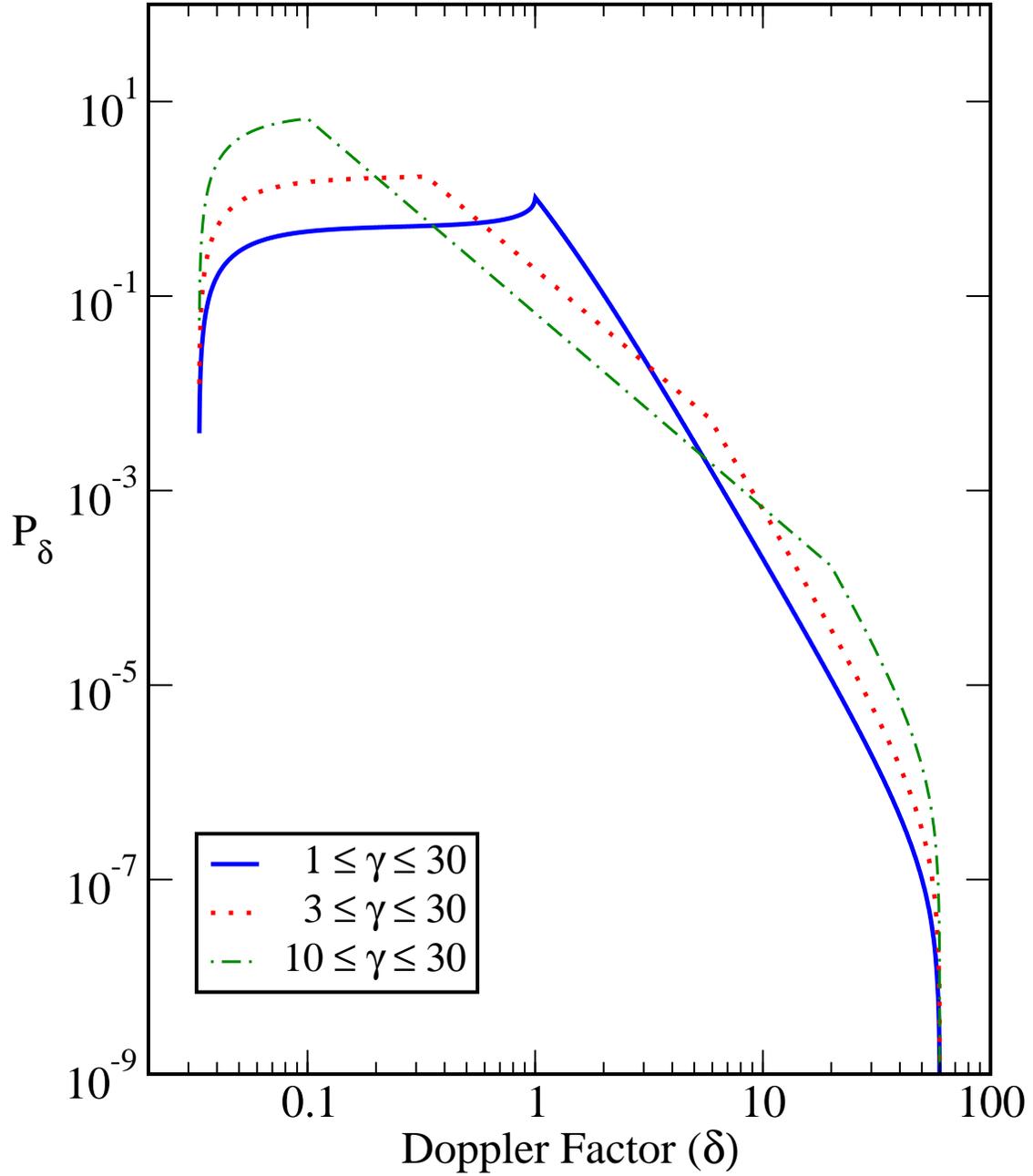}
\caption{\label{fi2} Predicted distribution functions of Doppler factor
for jet populations with bulk Lorentz factors distributed according to
$P_{\gamma} \propto \gamma^{-2}$, and ranging up to $\gamma_2 = 30$. The
solid curve shows the predicted distribution for a lower limits of
$\gamma_1 =1$, while the dotted and dot-dashed curves represent models
with $\gamma_1 = 3$ and $\gamma_1 =10$, respectively.  }
\end{figure*}

We have shown that the power-law slope of the Lorentz factor
distribution can have a strong influence on the distribution of Doppler
factors in a randomly-oriented jet population. In Figure~\ref{fi2} we
show the effects of increasing the lower limit of the Lorentz factor
distribution on a population with $\gamma_1 \le \gamma \le \gamma_2$
and $P_{\gamma} \propto \gamma^{-2}$.  As
$\gamma_1$ is increased, the region $\gamma_1^{-1} <
\delta < \gamma_1(1+\beta_1)$ grows wider, and the cusp at $\delta = \gamma_1^{-1}$
 is replaced by a broad region with slope equal to $-2$. The
same behavior also occurs for other values of $a$. Figure~\ref{fi2}
also confirms that the Doppler factor distribution approaches that of
the single-valued Lorentz factor case as $\gamma_1 \rightarrow
\gamma_2$.

It is important to note that in a sample of purely randomly-oriented
two-sided jets, exactly half will have approaching jet viewing angles
greater than $60\arcdeg$. This means that regardless of the form of
the jet Lorentz factor distribution, there will always be a
preponderance of sources with low Doppler factors, such that the mean
value of $\delta$ will always be less than unity. Somewhat
paradoxically, if a jet population is measured to have a high mean
Doppler factor, this actually implies that it must be dominated by {\it slow} 
jets. We caution, however, that this reasoning only applies to
genuinely orientation-unbiased samples. Flux-limited samples selected
on the basis of relativistically-beamed jet emission, on the other
hand, will be strongly biased toward small viewing angles, and as a
result will always be dominated by intrinsically fast jets with high
Doppler factors (i.e., blazars; see \citealt*{VC94}).


\section{BEAMED LUMINOSITY FUNCTIONS \label{beamedlfsec}}
We now use our results on the predicted Doppler factor distributions
to derive the corresponding apparent luminosity functions of
relativistically beamed jet populations.  For the purposes of
obtaining analytical expressions, we will ignore any luminosity
contribution from the jet in each source that is pointing away from
the observer. In \S~\ref{twosidedsec} we will show using numerical
simulations that for a steep intrinsic LF, the receding (counter-jet)
luminosity only affects the shape of the faint end of the beamed LF,
and has no effect at high observed luminosities.

Due to relativistic aberration, the luminosity of an
isotropically-emitting, moving object according to a stationary
observer is
\begin{equation}L = \delta^p{\cal L}, \label{beameq} \end{equation}
where $\Ell$ is the source rest-frame luminosity, $p = 2 - \alpha$ for
continuous jet emission, and $\alpha$ is the spectral index ($S_\nu
\propto \nu^\alpha$). In the case of a discrete emitting region, we
have $p = 3 -\alpha$ due to relativistic time dilation of its finite
synchrotron-emitting lifetime \citep{LB85}. For fixed $p$, the
observed luminosity function of a population of relativistic jets will
depend solely on its Doppler factor distribution and its intrinsic
(unbeamed) luminosity function. In the figures that follow, we will
only plot models with $p = 2$, which is appropriate for 
the continuous jet emission associated with the flat-spectrum cores of
blazars.

We begin by assuming that the jet population has an intrinsic
power-law luminosity function of the form:
\begin{equation}\phi({\cal L}) = \cases{k_1 {\cal L}^{-B} & for
 ${\cal L}_1 \le {\cal L} \le {\cal L}_2$ \cr
0\; ,& elsewhere, } \end{equation} where the normalization constant is
\begin{equation}k_1 = {1-B \over {\cal L}_2^{1-B} - {\cal L}_1^{1-B}}
\; .\end{equation}
The transformation law for probability density functions yields
the joint probability function
\begin{equation}h(L,\delta) = P_{\delta}(\delta)\phi(\Ell)
 {d\Ell \over dL}.\label{hprob} \end{equation}
Using equations (\ref{beameq}) and (\ref{hprob}) and the intrinsic
luminosity function, we obtain
\begin{equation}h(L,\delta) = k_1  L^{-B}P_{\delta}(\delta)\;\delta^{-p(1-B)}.
\end{equation}
Integrating over $\delta$ gives
\begin{equation}\Phi(L) = k_1 L^{-B}
\int P_{\delta}(\delta)\;\delta^{Cp+1}d\delta, \label{phiL}\end{equation}
where (using the notation of \citealt*{US84}) 
\be C = B - (1/p) -1 \;. \label{Cdef} \ee

In their derivation of the beamed LF for a jet population having a
unique Lorentz factor, \cite{US84} identified several important values of
L. Here we keep their definitions of $L_3$ and $L_4$, and extend their
notation as follows:
\begin{eqnarray} L_{min} =   &\Ell_1 \gamma_2^{-p} \cr
L_5 = & \Ell_1\gamma_1^{-p} \cr
L_6 = & \Ell_1 (\gamma_1+\gamma_1\beta_1)^p \cr
L_4 = & \Ell_1 \;\delta_{max}^p \cr
L_3 = & \Ell_2\gamma_2^{-p} \cr
L_7 = &\Ell_2\gamma_1^{-p} \cr
L_8 = &\Ell_2(\gamma_1+\gamma_1\beta_1)^p \cr
L_{max} = &\Ell_2\;\delta_{max}^{p}.
\label{ldefs} 
\end{eqnarray}

We note that a similar extension of notation was made by \cite{UP91},
however, they do not include the parameter $L_5$, which is needed for the analytical solutions presented in Appendix B. 

When $\gamma_2$ is large, the beamed LF spans a wide range:
\be L_{max} / L_{min} = \gamma_2^{2p}(1+\beta_2)^p (\Ell_2/\Ell_1). \ee
Thus, even a jet population with a single intrinsic luminosity
$\Ell = \Ell_1 = \Ell_2$ will end up with a broad beamed luminosity
function. This scenario has been invoked to explain the wide range
of observed luminosities in gamma-ray bursts, which are thought to have
Lorentz factors on the order of 100 (e.g., \citealt*{KP00}).

To determine an analytical expression for the beamed LF, we substitute
the relevant expression for $P_{\delta}$ (see eq. \ref{unifpdelta})
into equation (\ref{phiL}) and integrate over appropriate limits.  We
find that analytical solutions are generally only possible when the
product $Cp$ is an integer. We derive expressions for several cases in
Appendix B. \ For simplicity, we have considered only cases where $B
\ge (p+1)/p$ and $L_3 > L_4$. This corresponds to the condition
$(\Ell_2 /
\Ell_1)^{1/p} > \gamma_2^2 (1+\beta_2)$, which will always be
satisfied provided $\Ell_2$ is made arbitrarily large. The latter can
be justified if, as in our case, the intrinsic LF slope is fairly steep (i.e.,
$B \ge 1.5$ for $p = 2$). With these choices of parameters, the $L$ values in
equation~(\ref{ldefs}) are arranged in order of increasing apparent
luminosity.

\begin{figure*}
\epsscale{0.9}
\plotone{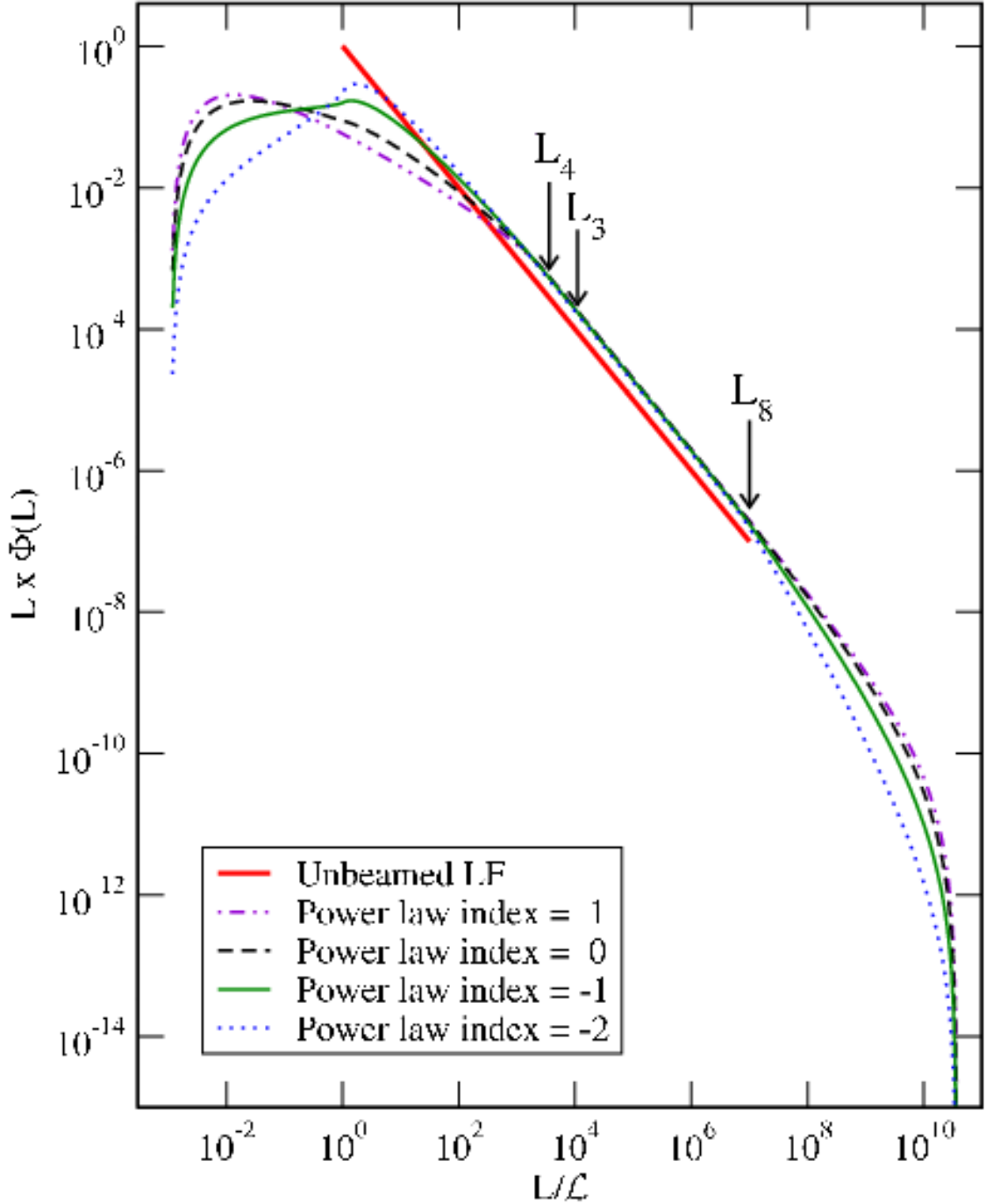}
\caption{\label{fi3} Predicted apparent luminosity functions (LFs) for jet
populations with Lorentz factors distributed according to a simple
power law. The intrinsic LF (thick solid curve) has a power-law
slope of $-2$, and ranges from $\Ell_1$ to $\Ell_2$, with $\Ell_2 /\Ell_1
= 10^7$.  The remaining curves represent the beamed (apparent) LFs for jet
populations with Lorenz factors distributed according to $P_{\gamma}
\propto \gamma^a$ over the range $1 \le \gamma \le 30$. These models
plotted correspond to $a = 1$ (thin solid curve), $a = 0$ (dotted
curve), $a = -1$ (dot-dashed curve), and $a = -2$ (dashed curve). The
specific locations of $L_3$, $L_4$, and $L_8$ are marked with arrows
(see eq.~\ref{ldefs}). }
\end{figure*}

In Figure~\ref{fi3} we plot $\log{\left[L\;\Phi(L)\right]}$ versus
$\log{[L/\Ell_1]}$ for various power-law Lorentz factor distributions
distributed over the range $1 \le \gamma \le 30$.  The thick solid
curve is the original (unbeamed) LF, which in this example has a
power-law slope equal to $-2$ and $\log{[\Ell_2 /\Ell_1]} = 7$. The
remaining curves represent the beamed LFs for various values of
power-law index $a$, where $P_{\gamma} \propto \gamma^a$.

\begin{figure*}   
\epsscale{0.9}
\plotone{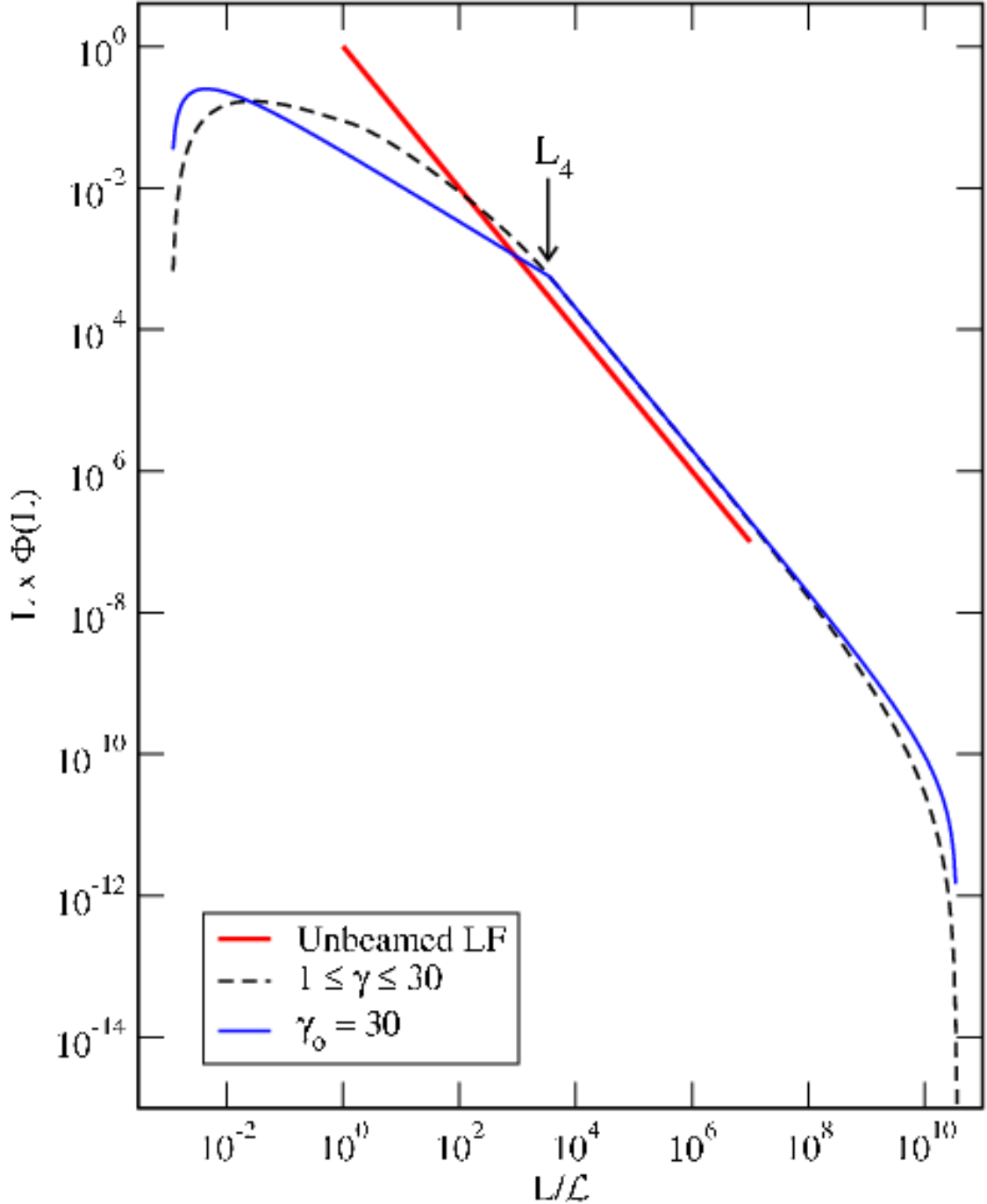}
\caption{\label{fi4} Comparison of beamed luminosity functions for jet
populations with single-valued and power-law Lorentz factor
distributions. The thick solid line represents the intrinsic LF, which
has a power-law slope of $-2$, and ranges over $\Ell_1$ to $\Ell_2$,
with $\Ell_2 /\Ell_1 = 10^7$.  The other curves represent beamed LFs
for a jet population with a single Lorentz factor $\gamma_o = 30$
(thin solid curve), one with Lorentz factors uniformly distributed
over the range $1 \le \gamma \le 30$ (dotted curve). }
\end{figure*}

\cite{US84} found that for a single-valued Lorentz factor distribution, the
beamed luminosity function will have the same slope as the unbeamed
one for $L > L_4$, except in the immediate vicinity of $L_{max}$. In
their numerical analysis, \cite{UP91} 
claimed that this also holds true for a distribution of Lorentz factors. We
confirm this in the case of flat or inverted Lorentz
factor distributions (i.e., $a = 0$ and $a = 1$). Indeed, the beamed
LFs of a single-valued model with $\gamma = \gamma_o$ and a uniformly
distributed model ($P_{\gamma} \propto \gamma\,;\; 1 \le \gamma \le
\gamma_o$) are nearly identical for $L > L_4$ (see Fig.~\ref{fi4}).

For steep power-law $\gamma$ distributions with $a \le -1$, the beamed
and unbeamed LFs also have the same slope, however, this is true only for the
region $L_4 < L < L_8$. Above $L_8$, the slope steepens by an amount
proportional to the power-law $\gamma$ index $a$. We note that in the
case of a jet population where $\Ell_2 / \Ell_1$ is large and the
intrinsic LF is steep, objects with $L > L_8$ will be extremely
rare. In the models shown in Figure 3, only one jet in $~\sim 10^7$
will have $L > L_8$.

At luminosities fainter than $L_4$, \cite{UP91} claimed that the slope
will always be equal to $-(p+1)/p$, regardless of the Lorentz factor
distribution. However, we find this is true only in cases where there
are very few low-Lorentz factor jets in the population (i.e., when
$\gamma_1 >> 1$ or $a > 0$). In cases where $\gamma_1 \simeq 1$ and
the Lorentz factor distribution is steep, the LF slope at low
luminosities is significantly flatter than $-(p+1)/p$, and can even be
highly inverted (see Fig~\ref{fi3}). We caution, however, that at
these low-luminosities many jets will lie in the plane of the sky, and
that our analytical expressions ignore the receding jet. Adding the
counterjet's contribution to the total emission will change the
predicted luminosity functions slightly. We discuss this issue in more
detail in the next section.
   
\clearpage

\clearpage
\begin{figure*}
\epsscale{0.9}
\plotone{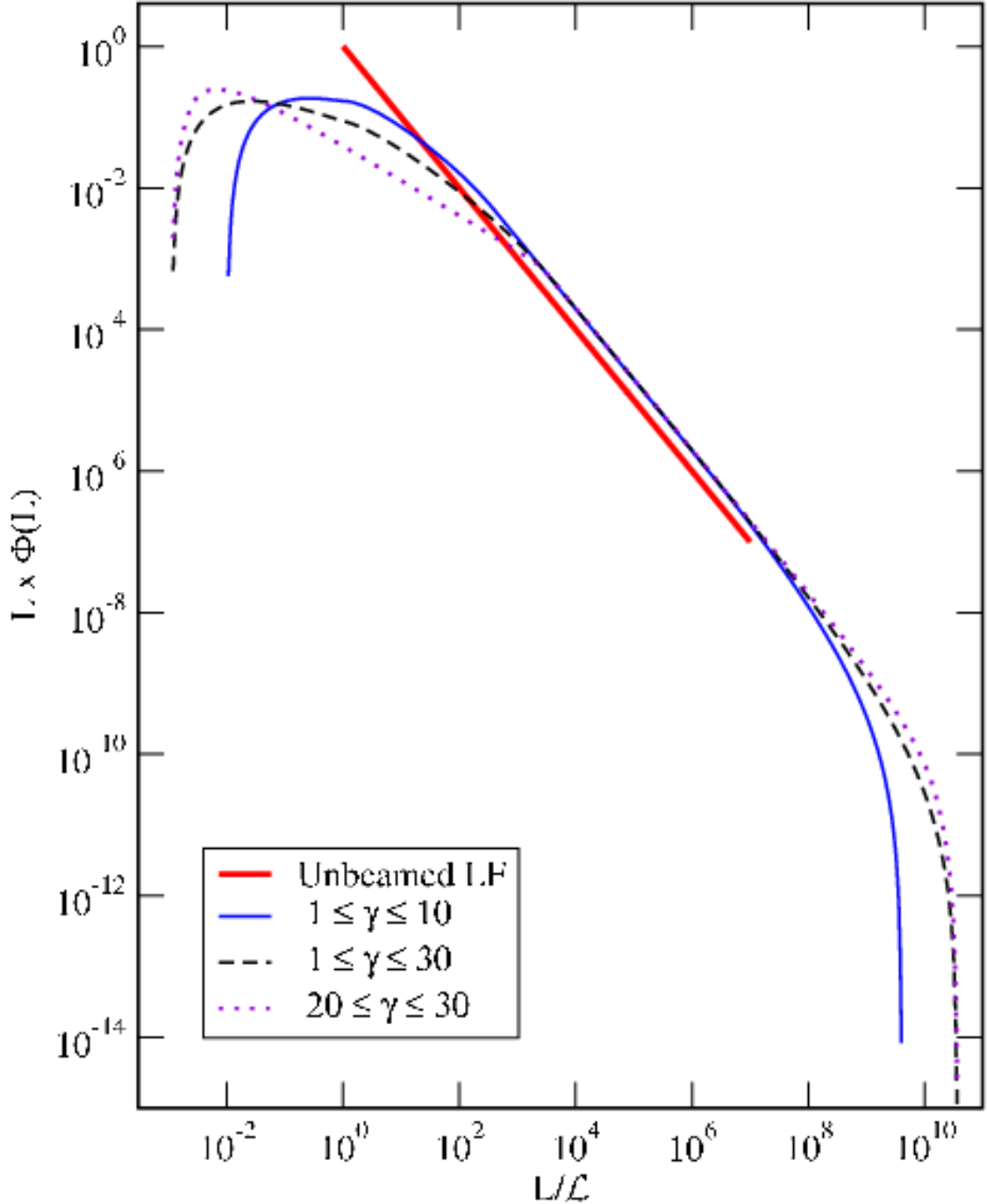}

\caption{\label{fi5}  Comparison of beamed luminosity functions for jet
populations with Lorentz factors uniformly distributed over different
ranges (i.e., $P_{\gamma} \propto$ constant, where $\gamma_1 \le \gamma
\le \gamma_2$).  The thick solid line represents the intrinsic LF,
which has a power-law slope of $-2$, and ranges over $\Ell_1$ to
$\Ell_2$, with $\Ell_2 /\Ell_1 = 10^7$.  The other curves represent
the beamed LFs of jet populations with $1 \le \gamma \le 10$ (thin
solid curve), $1 \le \gamma \le 30$ (dot-dashed curve), $20 \le \gamma
\le 30$ (dotted curve). }
\end{figure*}
\clearpage

\begin{figure*}
\epsscale{0.9}
\plotone{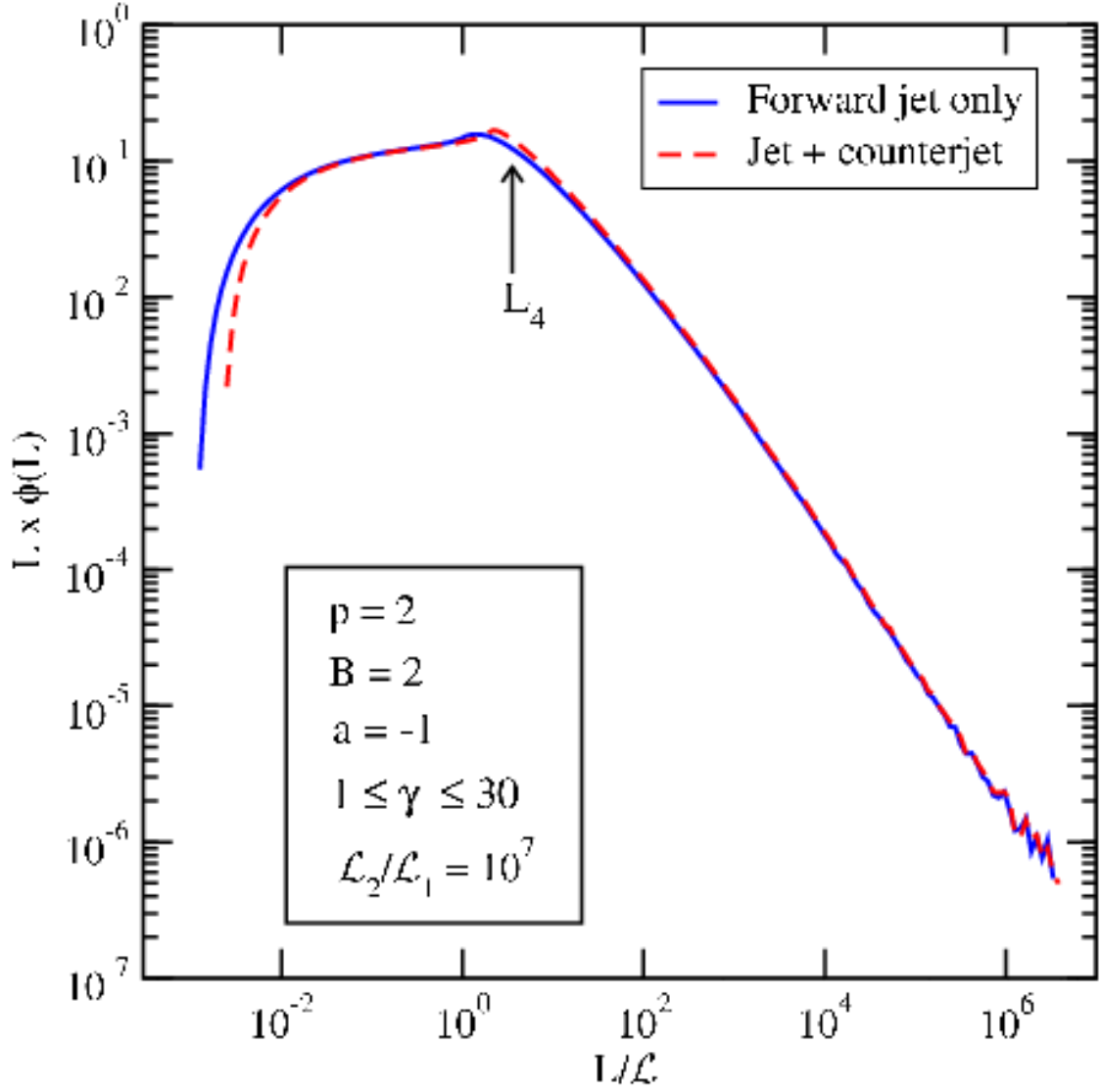}
\caption{\label{fi6} Beamed luminosity functions for two relativistic
two-sided jet populations with identical Lorentz factor distributions and
intrinsic luminosity functions. The dashed line is calculated
numerically, and shows the beamed LF if the luminosity from both the
approaching and receding jets in each source are considered. The solid
line shows the expected LF if the receding jet is ignored. The two curves
differ only at low apparent luminosities, where the Doppler factors
are small and the approaching and receding jets have similar luminosities. }
\end{figure*}


We have also investigated the effects of various ranges of Lorentz factor
on the apparent luminosity functions of jet populations.  In Figure~\ref{fi5} we plot
three different models with Lorentz factors uniformly distributed over
different ranges $\gamma_1 \le \gamma \le \gamma_2$. It is apparent
that for $L > L_4$, the bright end of the beamed LF is largely
insensitive to the values of $\gamma_1$ and $\gamma_2$. Increasing
$\gamma_2$ can potentially increase the number of ultra-high
luminosity objects, but again these are likely to be  extremely rare
for steep intrinsic LFs.

At faint luminosities, the beamed LF is highly sensitive to the value
of $\gamma_2$, since this dictates the minimum Doppler factor, and
hence $L_{min}$. Because a very fast jet population will have a low
mean Doppler factor (see \S~\ref{deltadiscussion}), its sources will,
on average, appear {\it fainter} than those of a population dominated
by slow speeds. Again, this reasoning applies only to genuinely
orientation-unbiased samples.

\subsection{Two-sided jets\label{twosidedsec}}

The beamed luminosity functions presented in the previous section are
not strictly correct at low luminosities, since they ignore any
luminosity contribution from the receding jet in each source. When
considering only highly beamed objects such as blazars or gamma-ray
bursts, this is a reasonable approximation, since the receding jet
will be extremely faint (and usually undetectable). However, in
samples containing mildly relativistic jets, or jets that lie close to
the plane of the sky (e.g., lobe-dominated radio galaxies), both the
approaching and receding jets can contribute roughly equal amounts of
flux.

In Figure~\ref{fi6} we compare two models with $B = 2$, $a = -1$, $1
\le \gamma \le 30$, and $\Ell_2 /\Ell_1 = 10^7$. The solid line shows the
beamed luminosity function if only the luminosity of the approaching jet
is considered. The dashed line represents a numerically-determined
beamed LF in which the luminosities of both the approaching and receding
jets are taken into account. For luminosities below $L_1$, the curves
differ considerably, since these jets lie close to the plane of the
sky. At higher luminosities, the predicted LFs are virtually
identical, since the apparent approaching/receding jet luminosity ratio
becomes very large for small viewing angles and high Lorentz
factors. We find that this result holds true for all ranges of $B$ and
$a$ discussed in this paper. We conclude therefore, that the
emission from the receding jet can be safely ignored when studying the
luminosity functions of highly beamed objects.

\section{OBSERVATIONAL IMPLICATIONS\label{obsconseq}}

The analytical results derived in this paper are applicable to any
astronomical sample that is selected solely on the basis of
relativistically-beamed jet emission. Our results are particularly
relevant to compact radio-loud AGN, whose high-frequency radio
emission originates primarily from milliarcsecond-scale relativistic
jets. With the advent of dedicated VLBI arrays, it is now possible to
conduct large multi-epoch surveys of these objects (e.g., the
Caltech-Jodrell Surveys; \citealt*{TRP96} and the VLBA 2 cm Survey;
\citealt*{KVZ98}). However, because of the high thresholds required
for fringe detection, these large VLBI studies have so far been
confined to the brightest, most compact objects whose radiation is
highly beamed toward us. These types of samples are not ideal for
studying the general properties of the AGN jets, since one must first
disentangle the combined effects of beaming, jet axis orientation
bias, and Malmquist bias before one can construct a luminosity
function or intrinsic speed distribution. Monte Carlo simulations have
played an important role in this respect. \cite{LM97} showed that
based on their observed superluminal speed distribution, AGN jets must
have a wide range of Lorentz factors distributed according to a
relatively steep power law ($P_{\gamma} \propto \gamma^a$, where $a
\simeq -1.5$).  They also demonstrated that a reasonably large, flux-limited sample of
bright jets should always contain some objects with apparent speed
equal to the maximum jet Lorentz factor in the parent population.

Since the observed LF of radio-loud AGN is fairly steep $(B \simeq 1.9
\pm 0.1$; \citealt*{WRB98}), our findings suggest that the observed LF
of a complete flux-limited sample selected on the basis of beamed
emission will closely match the intrinsic LF for luminosities in the
range $4\Ell_1\beta_{a,max}^2 \lesssim L \lesssim \Ell_2$. Here we
have assumed a boost index $p = 2$ and $\gamma_2$ equal to the maximum
apparent speed $\beta_{a,max}$. The lower luminosity limit $\Ell_1$
can be constrained using the nearest sources in a complete VLBI flux-limited
sample, since these will generally meet the flux density criteria on
the basis of their cosmological proximity, and not their Doppler boosting
factors. The upper luminosity limit $\Ell_2$ can be constrained using
the observed space density of radio-loud AGN. 

We are currently gathering VLBI proper motion and flux density data at
15 GHz on a sample of the brightest $\sim 130$ jets in the northern
sky. This sample, entitled MOJAVE (Monitoring Of Jets in AGN with VLBA
Experiments; \citealt*{Z03}), is complete on the basis of compact
(VLBI) flux density, and is therefore well-suited for exploring the
intrinsic luminosity function of AGN jets. An analysis of the MOJAVE
data is currently underway, and the results will be discussed in an
upcoming paper.

\section{SUMMARY \label{summary}}

Astronomical samples of objects selected on the basis of emission from
relativistic jets, such as gamma-ray bursters, gamma-ray loud AGN, and
compact radio-loud AGN, will by highly affected by relativistic
beaming effects. In particular, their observed luminosities are
predicted to be enhanced by a factor $\delta^p$, where $\delta$ is the
Doppler factor. We have
derived expressions for the predicted Doppler factor distributions of
two-sided, randomly-oriented, straight jet populations whose bulk
Lorentz factors are distributed in the range $[\gamma_1, \gamma_2]$
according to a simple power-law $P_{\gamma}
\propto \gamma^a$. We summarize our findings as follows:
\begin{itemize}

\item The  minimum and maximum jet Doppler factor in the population
will depend only on the maximum Lorentz factor
$\gamma_2$. Specifically, $\delta_{min} =
\gamma_2^{-1}$, and $\delta_{max} = \gamma_2(1+\beta_2)$.


\item An orientation-unbiased sample of jets will always have a mean
Doppler factor less than unity, since the majority of jets will lie
close to the plane of the sky. Accordingly, a jet population weighted
toward very fast speeds will have a lower mean Doppler factor, and
therefore appear fainter than an identical jet population with slower
intrinsic jet speeds. 

\item A randomly-oriented jet population with a flat ($a = 0$) or inverted ($a = 1$)
Lorentz factor distribution will have a Doppler factor distribution
very similar to a population where all jets have $\gamma =
\gamma_2$. Over a broad range of $\delta$, the distribution function
$P_{\delta}$ will vary as $\delta^{-2}$, except near $\delta_{min}$ and
$\delta_{max}$, where much fewer sources are expected compared to the
single-valued case. Sources with the highest possible Doppler factors
near $\delta_{max}$ will always be rare, even when all the jets in the
population are highly relativistic (i.e., $\gamma_1 >> 1$). This is a
consequence of the extremely small angles to the line of sight that are
necessary for high Doppler factors.
\end{itemize}

We have also derived analytical expressions for the apparent 
luminosity functions (LFs) of relativistic jet populations having a power-law
distribution of Lorentz factors and a steep intrinsic power-law luminosity
function. We find that:
\begin{itemize}
\item The slope of the apparent LF at low luminosities is not always
equal $-(p+1)/p$, as claimed by \cite{UP91}. The apparent LF in this
region is highly sensitive to the speed distribution of the parent
population. In particular, if the parent population is dominated by
slow jets, the LF slope will be very flat or inverted at low
luminosities.

\item Regardless of the values of $\gamma_1$ and $\gamma_2$, when the
Lorentz factor distribution of the jet population is steep (i.e., $a
\le -1$, where   $P_{\gamma} \propto \gamma^a$), the intrinsic and
beamed LFs will  have the same slope over the range $\Ell_1
\delta_{max}^p \le L \le \Ell_2$, where $L$ is the observed luminosity and $\Ell_1$ and $\Ell_2$ are the lower and  upper cutoffs
on the intrinsic LF, respectively.

\item Since the intrinsic LF of radio-loud AGN jets is known to be
steep, it should be possible to investigate its properties using
flux-limited VLBI samples of highly beamed AGN jets. Such samples will
have to be selected at relatively high radio frequencies (e.g. above
$\sim 10$ GHz) in order to eliminate any contribution from un-beamed,
steep-spectrum emission. An analysis of such a sample selected at 15
GHz (MOJAVE) is underway, and will be presented in an upcoming paper.

\end{itemize}

\acknowledgements

The author wishes to thank Ken Kellermann, Marshall Cohen, Tigran
Arshakian, and Rene Vermeulen for helpful comments on this manuscript.

\begin{appendix}

\section{DOPPLER FACTOR DISTRIBUTIONS OF JET POPULATIONS WITH A POWER
LAW DISTRIBUTION OF LORENTZ FACTORS}

\begin{figure*}
\epsscale{1.0}
\plotone{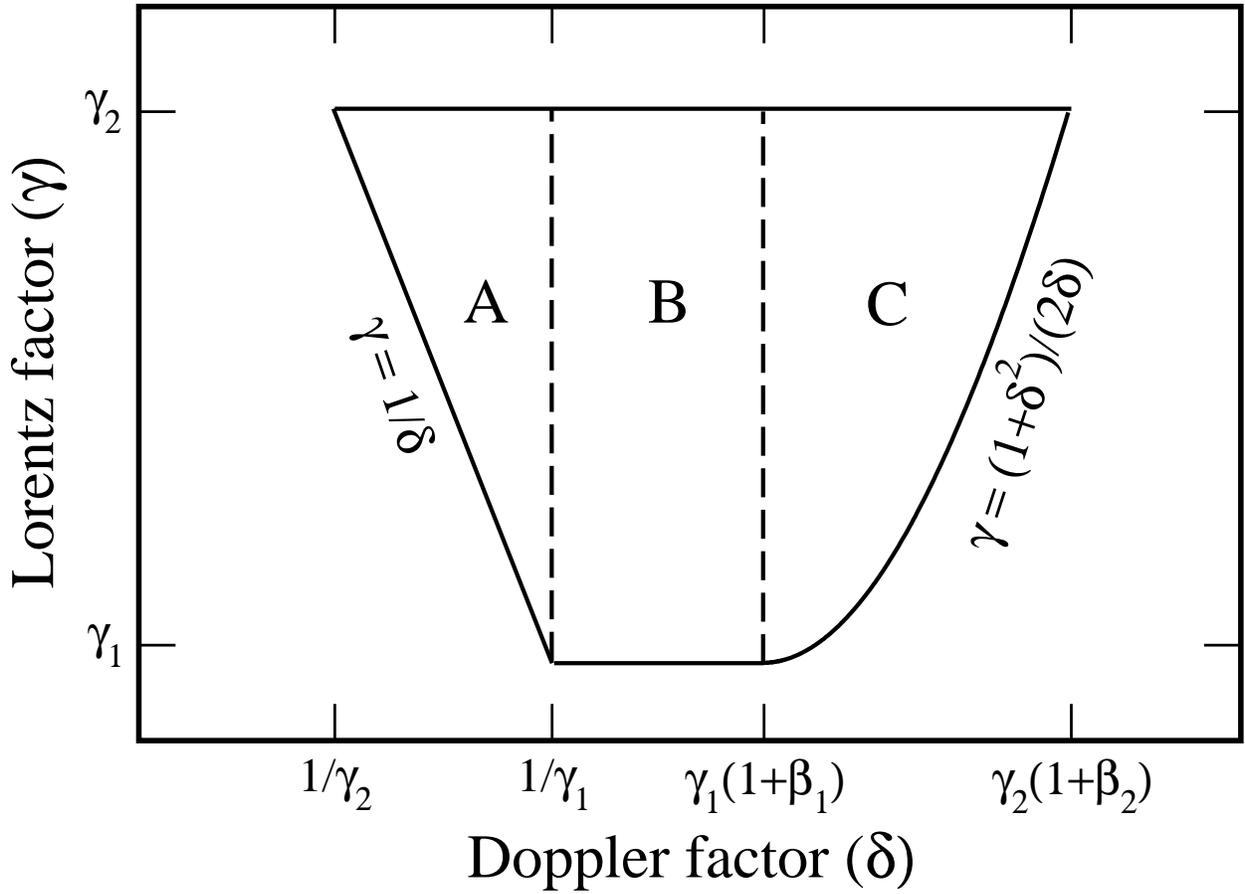}
\caption{\label{fi7} Plot of Lorentz factor versus Doppler factor
illustrating integration limits for determining $P_{\delta}$ (see
equation~\ref{pdeltaeq}). For a particular value of $\delta$, the
integration limits on $\gamma$ are defined by the polygonal region (i.e., the
lower integration limit is equal to $\delta^{-1}$ in region A,
$\gamma_1$ in region B, and $(1+\delta^2)/2\delta$ in region C.) }
\end{figure*}

We wish to determine the probability density function $P_{\delta}(\delta)$,
which describes the expected distribution of Doppler factors for a
randomly-oriented, two-sided jet population. The theory of probability
transformation for several variables implies
\begin{equation} g(\delta, \gamma) = P_{\gamma}(\gamma) P_{\theta}(\theta) \left|d\theta \over
d\delta\right|.  \end{equation}
Now
\begin{equation}\left|{d\delta \over d\theta}\right| = \delta^2\sqrt{\gamma^2-1}\;\sin{\theta}.  \end{equation}
Since the viewing angles are distributed according to 
\begin{equation} P_{\theta}(\theta) = \sin{\theta}, \end{equation}
we have
\begin{equation} g(\delta, \gamma) = {
P_{\gamma}(\gamma) \over \delta^2  \sqrt{\gamma^2-1}\;}. \end{equation}
We integrate over $\gamma$ to obtain the probability density function
for $\delta$:
\begin{equation}P_{\delta}(\delta) = \delta^{-2} \int_{f(\delta)}^{\gamma_{2}}
{P_{\gamma}(\gamma) \over  \sqrt{\gamma^2-1}}\;d\gamma, \label{pdeltaeq}  \end{equation}
where the limits on this integral are illustrated graphically in
Figure~\ref{fi7}. The lower limit depends on the value of $\delta$ as follows:  
\begin{equation}f(\delta) = \cases{\delta^{-1} & for
$\quad \gamma_{2}^{-1} \le \delta < \gamma_{1}^{-1}$ \cr
\gamma_1  & for
$\quad \gamma_{1}^{-1} \le \delta < \gamma_1(1+\beta_1)$ \cr
{(1+\delta^2) \over 2\delta}  & for $\quad \gamma_1(1+\beta_1)  \le 
\delta \le \delta_{max}$ \cr
\gamma_2 & elsewhere, \cr}  \end{equation}
where we have used the fact that $\sqrt{\gamma^2-1}= \gamma\beta$ and
$\delta_{max} = \gamma_2(1+\beta_2)$.

Analytical solutions of equation~(\ref{pdeltaeq}) are generally
possible if $P_{\gamma} \propto \gamma^a$, where $a$ is an
integer. Below we derive expressions for $P_{\delta}(\delta)$ for power-law
slopes ranging from $a = +1$ to $a = -2$.

\subsection{Case 1: Single-valued Lorentz factor distribution}

When $\gamma_1 = \gamma_2 = \gamma_o$, then $P_{\gamma}(\gamma)$
can be considered as a delta-function centered at $\gamma = \gamma_o$. Equation~(\ref{pdeltaeq}) yields the well-known result (e.g., \citealt*{US84}):

\begin{equation} P_{\delta}(\delta) = \cases{  \delta^{-2} \;\gamma_o^{-1}\; \beta_o^{-1} & for $
\gamma_o^{-1} \le \delta \le \gamma_o(1+\beta_o)$ \cr
0\; ,& elsewhere. } \label{singleval} \end{equation}

\subsection{ Case 2: Uniform Lorentz factor distribution \label{uniformsec}}

In the case where the Lorentz factors in the jet population are
uniformly distributed within the range $\gamma_1 \le \gamma \le
\gamma_2$, this is equivalent to setting $a = 0$, i.e.:
\begin{equation}P_{\gamma}(\gamma) = \cases{\left(\gamma_{2} -
\gamma_{1}\right)^{-1} &for $\gamma_{1} \le \gamma \le
\gamma_{2}$\cr
 0& elsewhere,} \end{equation}

we substitute this function into equation~(\ref{pdeltaeq})  and perform the
integration to obtain
\begin{equation}P_{\delta}(\delta) ={ 1 \over \delta^2  \left(\gamma_{2} - \gamma_{1}\right)} \times
 \cases{  \ln{\left|\delta \delta_{max} \over 1 + \sqrt{1-\delta^2}\right|} & for $\quad \gamma_{2}^{-1} \le \delta < \gamma_{1}^{-1}$ \cr   \ln{\left|{\delta_{max} \over \gamma_1(1+\beta_1)}\right|}  & for  $\quad \gamma_{1}^{-1} \le \delta < \gamma_1(1+\beta_1)$ \cr 
  \ln{\left|\delta_{max} /\delta\right|} & for  $\quad \gamma_1(1+\beta_1) \le \delta \le \delta_{max}$ \cr
0\; ,& elsewhere.} \label{unifpdelta} \end{equation}

For large   $\gamma_{2}$ and $\gamma_{1} \simeq 1$, equation~(\ref{unifpdelta}) reduces to:
\begin{equation}P_{\delta}(\delta) \simeq {1 \over \delta^2\gamma_{2}} \times \cases{\ln{\left|{2 \delta \gamma_2 \over 1 + \sqrt{1 - \delta^2}}\right|} & for $\quad \gamma_{2}^{-1} \le \delta < 1$ \cr
 \ln{\left|2\gamma_2 /\delta \right|}  & for $\quad 1 \le \delta < 2\gamma_2.$}\end{equation}

\subsection{Case 3: Power Law Lorentz Factor Distributions}

\subsubsection{Power law distribution of the form  $P_{\gamma}(\gamma) \propto
\gamma$}
For the case $a = +1$ we have 
\begin{equation}
P_{\gamma}(\gamma) = \cases{2\left(\gamma_2^2 - \gamma_1^2\right)^{-1} &for $\gamma_{1} \le \gamma \le \gamma_{2}$\cr
 0 &elsewhere.}
\end{equation}

Substituting this into equation~(\ref{pdeltaeq}) gives
\begin{equation} P_{\delta}(\delta; a = +1) = { 2 \over \delta^2
\left(\gamma_2^2 - \gamma_1^2\right)  }\times \cases{\gamma_2\beta_2 -
{\sqrt{1-\delta^2} \over \delta}&  for $\quad \gamma_{2}^{-1} \le \delta < \gamma_1^{-1}$ \cr
 \gamma_2\beta_2-\gamma_1\beta_1 & for $\quad \gamma_{1}^{-1} \le \delta < \gamma_1(1+\beta_1)$ \cr 
\gamma_2\beta_2 - {\delta^2 -1 \over 2 \delta} & for $\quad \gamma_1(1+\beta_1) \le \delta \le \delta_{max}$\cr
0 & elsewhere.} \end{equation}

\subsubsection{Power law distribution of the form  $P_{\gamma}(\gamma) \propto
\gamma^{-1}$}
For the case $a = -1$ we have 
\begin{equation}
P_{\gamma}(\gamma) = \cases{\left(\gamma\ln{\left|{\gamma_2 / \gamma_1}\right|}\right)^{-1} &for $\gamma_{1} \le \gamma \le \gamma_{2}$\cr
 0 &elsewhere.}
\end{equation}

We obtain
\begin{equation} P_{\delta}(\delta; a = -1) = { 1 \over \delta^2 \ln{\left|{\gamma_2 / \gamma_1}\right|} }\times \cases{
\arctan{\left[\delta \over \sqrt{1-\delta^2}\right]} - \arctan{\left[1 \over \gamma_2\beta_2\right]} &  for $\quad \gamma_{2}^{-1} \le \delta < \gamma_1^{-1}$ \cr
 \arctan{\left[1 \over \gamma_1\beta_1\right]}- \arctan{\left[1 \over \gamma_2\beta_2\right]} & for $\quad \gamma_{1}^{-1} \le \delta < \gamma_1(1+\beta_1)$ \cr 
\arctan{\left[2\delta \over \delta^2-1\right]}- \arctan{\left[1 \over \gamma_2\beta_2\right]} & for $\quad \gamma_1(1+\beta_1) \le \delta \le \delta_{max}$\cr
0 & elsewhere.} \end{equation}

For  large $\gamma_{2} $ and $\gamma_{1} \simeq 1$, this reduces to:
\begin{equation} P_{\delta}(\delta; a = -1) = { 1 \over \delta^2 \ln{\left|{\gamma_2}\right|} }\times \cases{
\arctan{\left[\delta \over \sqrt{1-\delta^2}\right]}  &  for $\quad \gamma_{2}^{-1} \le \delta < 1$ \cr
\arctan{\left[2\delta \over \delta^2-1\right]} & for $\quad 1 \le \delta \le 2\gamma_2$.} \end{equation}

\subsubsection{Power law distribution of the form  $P_{\gamma}(\gamma) \propto \gamma^{-2}$}
For the case  $a = -2$, we have
\begin{equation}P_{\gamma}(\gamma) = \cases{\gamma^{-2}\left(\gamma_{1}^{-1} - \gamma_{2}^{-1}\right)^{-1} &for $\gamma_{1} \le \gamma \le \gamma_{2}$\cr 0 &elsewhere.}\end{equation}
We obtain
\begin{equation} P_{\delta}(\delta; a = -2) = { 1 \over \delta^2  \left(\gamma_{1}^{-1} -
\gamma_{2}^{-1}\right)}\times \cases{\beta_2 - \sqrt{1-\delta^2} & 
  for $\quad \gamma_{2}^{-1} \le \delta < \gamma_1^{-1}$ \cr
  \beta_2 - \beta_1 & for $\quad \gamma_{1}^{-1} \le \delta < \gamma_1(1+\beta_1)$ \cr 
 \beta_2 - {\delta^2-1 \over \delta^2+1} & for $\quad \gamma_1(1+\beta_1) \le \delta \le \delta_{max}$\cr
0 & elsewhere.} \end{equation}
   For large $\gamma_{2} $ and $\gamma_{1} \simeq 1$, this reduces to:
\begin{equation} P_{\delta}(\delta; a = -2) \simeq {1 \over \delta^2} \times \cases{ 1 -\sqrt{1-\delta^2} & for $\quad \gamma_{2}^{-1} \le \delta < 1$ \cr 
  2  \left(\delta^2+1\right)^{-1} & for $\quad 1 \le \delta < 2\gamma_2$.}\end{equation}

\section{BEAMED LUMINOSITY FUNCTIONS OF JET POPULATIONS WITH A POWER-LAW DISTRIBUTION OF LORENTZ FACTORS}

For $\gamma_1 \ne 1$ , the form of the beamed luminosity function is:
\be\scriptsize
\baselineskip=18pt
 \Phi(L) = {k_1 L^{-B}} \times \cases { 
0\; ,& $\quad L < L_{min}$ \cr
F[ (L/\Ell_1)^{1/p} ] - F[\gamma_2^{-1}], & $\quad L_{min} \le L < L_5$ \cr
G[(L/\Ell_1)^{1/p}] + F[ \gamma_1^{-1} ]- F[\gamma_2^{-1}]  - G[\gamma_1^{-1}], & $\quad L_5 \le L  < L_6$  \cr
H[(L/\Ell_1)^{1/p}] + F[\gamma_1^{-1}]-F[\gamma_2^{-1}] +G[\gamma_1 + \gamma_1 \beta_1] -G[\gamma_1^{-1}]  - H[\gamma_1 + \gamma_1 \beta_1], & $\quad L_6 \le L < L_4$ \cr
F[\gamma_1^{-1}]-F[\gamma_2^{-1}]+G[\gamma_1 + \gamma_1 \beta_1] -G[\gamma_1^{-1}] + H[\delta_{max}] - H[\gamma_1 + \gamma_1 \beta_1], & $\quad L_4 \le L <  L_3$ \cr
-F[(L/\Ell_2)^{1/p}] + F[\gamma_1^{-1}]  +G[\gamma_1 + \gamma_1 \beta_1] -G[\gamma_1^{-1}] + H[\delta_{max}] - H[\gamma_1 + \gamma_1 \beta_1] , & $\quad L_3 \le L< L_7$\cr
-G[(L/\Ell_2)^{1/p}] + G[\gamma_1 + \gamma_1 \beta_1] + H[\delta_{max}] - H[\gamma_1 + \gamma_1 \beta_1], & $\quad L_7 \le L < L_8$\cr
- H[(L/\Ell_2)^{1/p}] +H[\delta_{max}]  , & $\quad L_8\le L \le L_{max}$\cr
0\; ,& $\quad L > L_{max}$. } \label{PhiLunif}
\ee

If $\gamma_1 = 1$, then $L_5 = L_6 = \Ell_1$, and $L_7 = L_8 =
\Ell_2$, and equation~(\ref{PhiLunif}) reduces to:
\be \Phi(L) = k_1  L^{-B}\times \cases { 
0\; ,& $\quad L < L_{min}$ \cr
F[ (L/\Ell_1)^{1/p} ] - F[\gamma_2^{-1}], & $\quad L_{min} \le L < \Ell_1$ \cr
 H[(L/\Ell_1)^{1/p}] -H[1]+ F[1]-F[\gamma_2^{-1}]   & $\quad \Ell_1 \le L < L_4$ \cr
F[1]-F[\gamma_2^{-1}] + H[\delta_{max}] - H[1],& $\quad L_4 \le L <  L_3$ \cr
-F[(L/\Ell_2)^{1/p}] + F[1] + H[\delta_{max}] - H[1], & $\quad L_3 \le L< \Ell_2$\cr
-H[(L/\Ell_2)^{1/p}] + H[\delta_{max}], & $\quad \Ell_2 \le L \le L_{max}$\cr
0\; ,& $\quad L > L_{max}$ \cr} \label{PhiLunif2}
\ee

The functions $F(x)$, $G(x)$, and $H(x)$ are tabulated in the
following sections for various integer values of $a$ and $Cp$. They
are grouped by particular values of $a$, where $P_{\gamma}(\gamma)
\propto \gamma^a$ in the jet population, and are further subdivided by
particular values of the product $Cp$ (see eq.~\ref{Cdef}).

\subsection{Power-Law Lorentz Factor Distribution of the
form $P_{\gamma}(\gamma) \propto \gamma $}

When $a = +1$ we have:
\begin{eqnarray} F(x) =
{ 2\over \gamma_2^2-\gamma_1^2} \times \cases{ 
\gamma_2\beta_2 \ln{|x|} + \arcsin{x} + {\sqrt{1-x^2} \over x} , & if $\quad Cp = 0$ \cr
 \gamma_2\beta_2 x -\sqrt{1-x^2} - \ln{\left|{x \over 1 + \sqrt{1-x^2}}\right|} , & if $\quad Cp = 1$ \cr
{1 \over 2} \gamma_2\beta_2 x^2 - {1\over 2} \arcsin{x} - {x \over 2} 
\sqrt{1-x^2},& if $\quad Cp = 2$ \cr
{1 \over 3} \gamma_2\beta_2 x^3 +{1 \over 3}(1-x^2)^{3/2},& if $\quad Cp = 3$,}
\end{eqnarray}

\begin{eqnarray} G(x) = 
{2 \left(\gamma_2 \beta_2 - \gamma_1 \beta_1
\right) \over \left(\gamma_2^2-\gamma_1^2\right)} \times \cases{
\ln{|x|}, & if $\quad Cp = 0$ \cr
{x^{Cp} \over Cp}, & if $\quad Cp = 1,2, 3$,}
\end{eqnarray}

\begin{eqnarray} H(x) =
{ 2 \over \gamma_2^2 - \gamma_1^2} \times \cases{ 
\gamma_2\beta_2 \ln{|x|} - {x \over 2} - {1 \over 2 x} , & if $\quad
Cp = 0$ \cr
\gamma_2\beta_2 x - {x^2 \over 4} - {1 \over 2 }\ln{|x|} , & if $\quad
Cp = 1$ \cr
{\gamma_2\beta_2 x^{Cp} \over Cp} - {x^{Cp+1} \over 2(Cp+1)}
- {x^{Cp-1} \over 2(Cp-1)},& if $\quad Cp = 2,3$. }
\end{eqnarray}

\subsection{Uniform Lorentz factor distribution of the form $P_{\gamma}(\gamma)= constant$}
In the case of a uniform $\gamma$ distribution ($a = 0$), we have 
\begin{eqnarray} F(x) = {1 \over \gamma_2 - \gamma_1} \times \cases{ x \ln{\left|{x\; \delta_{max} \over 1 + \sqrt{1-x^2}}\right|} - \arcsin{ x}, & if $\quad Cp = 1$ \cr
{1\over 2}\sqrt{1-x^2} + {x^2 \over 2} \ln{\left|{x \delta_{max} \over 1 + \sqrt{1-x^2}}\right|}, & if $\quad Cp = 2$ \cr
{x \over 6}\sqrt{1-x^2} -{1\over 6} \arcsin{x}+ {x^3 \over 3} \ln{\left|{x \delta_{max} \over 1 + \sqrt{1-x^2}}\right|} & if $\quad Cp = 3$, }
\end{eqnarray}
\be G(x) =  { x^{Cp} \ln{\left|\delta_{max} \over \gamma_1(1+\beta_1)\right|} \over (\gamma_2 - \gamma_1)Cp} , \quad {\rm if}\quad Cp = 1,2,3, \ee
\be H(x) =  { x^{Cp}\left(1+ Cp\;\ln{\left|\delta_{max}/x\right|}\right) \over(\gamma_2 - \gamma_1) (Cp)^2 } \quad \quad {\rm if} \quad Cp = 1,2,3. \ee

There are no analytical expressions possible when $a = 0$ and $Cp = 0$. 

\subsection{Power-Law Lorentz Factor Distribution of the
form $P_{\gamma}(\gamma) \propto \gamma^{-1}$}

When $a = -1$, we have 
\begin{eqnarray} F(x) =
{ 1 \over \ln{\left|\gamma_2/\gamma_1\right|}} \times \cases{ 
x\left[\arctan{\left(x \over \sqrt{1-x^2}\right)}-\arctan{\left({1 \over \gamma_2\beta_2}\right)}\right]+ \sqrt{1-x^2} , & if $\quad Cp = 1$ \cr
{x^2 \over 2}\left[\arctan{\left(x \over \sqrt{1-x^2}\right)} -\arctan{\left({1 \over \gamma_2\beta_2}\right)}\right] + {x\over 4}\sqrt{1-x^2}-{1 \over 4}\arcsin{[x]}  , & if $\quad Cp = 2$ \cr
{x^3 \over 3}\left[\arctan{\left(x \over \sqrt{1-x^2}\right)} -\arctan{\left({1 \over \gamma_2\beta_2}\right)}\right]+{1\over 9}(x^2+2)\sqrt{1-x^2} ,& if $\quad Cp = 3$, }
\end{eqnarray}

\be G(x) = {x^{Cp} \over Cp\ln{\left|\gamma_2/\gamma_1\right|}}
\left[\arctan{\left({1 \over \gamma_1\beta_1}\right)}-\arctan{\left({1
\over \gamma_2\beta_2}\right)}\right] , \quad {\rm if}\quad Cp = 1,2,3, \ee

\begin{eqnarray} H(x) =
{ 1 \over \ln{\left|\gamma_2/\gamma_1\right|}} \times \cases{ 
x\left[\arctan{\left(2x \over x^2-1\right)} -\arctan{\left({1 \over \gamma_2\beta_2}\right)}\right] + \ln{\left|x^2+1\right|} , & if $\quad Cp = 1$ \cr
{x^2 \over 2}\left[\arctan{\left(2x \over x^2-1\right)} -\arctan{\left({1 \over \gamma_2\beta_2}\right)}\right] +x -\arctan{x}   , & if $\quad Cp = 2$ \cr
{x^3 \over 3}\left[\arctan{\left(2x \over x^2-1\right)} -\arctan{\left({1 \over \gamma_2\beta_2}\right)}\right]+ {x^2 \over 3}-{1 \over 3}\ln{\left|x^2+1\right|} ,& if $\quad Cp = 3$. }
\end{eqnarray}

There are no analytical expressions possible when $a = -1$ and $Cp = 0$. 
\subsection{Power-Law Lorentz Factor Distribution of the
form $P_{\gamma}(\gamma) \propto \gamma^{-2}$}

For the case $a = -2$, we have
\begin{eqnarray} F(x) =
{ 1 \over \gamma_1^{-1}-\gamma_2^{-1}} \times \cases{ 
(\beta_2 -1)\ln{x}-\sqrt{1-x^2} + \ln{\left|1+\sqrt{1-x^2}\right|} , & if $\quad Cp = 0$ \cr
\beta_2 x - {x\over 2}\sqrt{1-x^2} -{1\over 2} \arcsin{x}, & if $\quad Cp = 1$ \cr
{\beta_2 x^2 \over 2} +{1 \over 3}(1-x^2)^{3/2} , & if $\quad Cp = 2$ \cr
{\beta_2 x^3 \over 3}-{1\over 8} \arcsin{x} - {x \over 8}(2x^2-1)\sqrt{1-x^2},& if $\quad Cp = 3$, }
\end{eqnarray}

\begin{eqnarray} G(x) = 
{(\beta_2 - \beta_1)  \over \left(\gamma_1^{-1}-\gamma_2^{-1}\right)} \times \cases{
\ln{|x|}, & if $\quad Cp = 0$ \cr
{x^{Cp} \over Cp}, & if $\quad Cp = 1,2, 3$,}
\end{eqnarray}

\begin{eqnarray}
H(x) = { 1 \over \gamma_1^{-1}-\gamma_2^{-1}} \times  \cases{ 
(\beta_2 + 1) \ln{x} - \ln{\left|1+x^2\right|}, & if $\quad Cp = 0$ \cr
(\beta_2 -1) x + 2 \arctan{x}, & if $\quad Cp = 1$ \cr
{1 \over 2}(\beta_2 - 1) x^2   + \ln{\left|1 + x^2\right|}, & if $\quad Cp = 2$ \cr
{1 \over 3}(\beta_2 - 1)x^3+2x-2 \arctan{x}, & if $\quad Cp = 3$.}
\end{eqnarray}
\end{appendix}


\begin{thebibliography}{}
\bibitem[Jackson \& Wall(1999)]{JW99} Jackson, C.~A.~\& 
Wall, J.~V.\ 1999, \mnras, 304, 160 

\bibitem[Kellermann et al.(1998)]{KVZ98} Kellermann, K.~I., Vermeulen, R.~C., 
Zensus, J.~A., \& Cohen, M.~H.\ 1998, \aj, 115, 1295 

\bibitem[Kellermann et al.(2003)]{K03} Kellermann, K. I. et al, 2003,
in preparation

\bibitem[Kumar \& Piran(2000)]{KP00} Kumar, P.~\& Piran, T.\ 
2000, \apj, 535, 152 

\bibitem[Lind \& Blandford(1985)]{LB85} Lind, K.~R.~\& 
Blandford, R.~D.\ 1985, \apj, 295, 358 

\bibitem[Lister \& Marscher(1997)]{LM97} Lister, M.~L.~\& 
Marscher, A.~P.\ 1997, \apj, 476, 572 


\bibitem[Taylor et al.(1996)]{TRP96} Taylor, G.~B., 
Vermeulen, R.~C., Readhead, A.~C.~S., Pearson, T.~J., Henstock, D.~R., \& 
Wilkinson, P.~N.\ 1996, \apjs, 107, 37 

\bibitem[Urry \& Padovani(1995)]{UP95} Urry, C.~M.~\& 
Padovani, P.\ 1995, \pasp, 107, 803 

\bibitem[Urry \& Padovani(1991)]{UP91} Urry, C.~M.~\& 
Padovani, P.\ 1991, \apj, 371, 60 

\bibitem[Urry, Padovani, \& Stickel(1991)Urry et al.]{UPS91} Urry, 
C.~M., Padovani, P., \& Stickel, M.\ 1991, \apj, 382, 501

\bibitem[Urry \& Shafer(1984)]{US84} Urry, C.~M.~\& Shafer, 
R.~A.\ 1984, \apj, 280, 569 (US84)

\bibitem[Vermeulen \& Cohen(1994)]{VC94} Vermeulen, R.~C.~\& 
Cohen, M.~H.\ 1994, \apj, 430, 467 

\bibitem[Vermeulen et al.(2003)]{V03} Vermeulen, R. C., \& Britzen,
S. 2003, in Radio Astronomy at the Fringe, Proceedings of a Workshop on Extragalactic Radio Sources held in Green Bank, WV, Oct 10-12, 2002, eds. J.A. Zensus, E. Ros, \& M.H.C. Cohen, in press.

\bibitem[Willott et al.(1998)]{WRB98} Willott, C.~J., Rawlings, S.,
Blundell, K.~M., \& Lacy, M.\ 1998, \mnras,  300, 625 


\bibitem[Zensus et al.(2003)]{Z03} Zensus, J. A., et al. 2003, in
Radio Astronomy at the Fringe, Proceedings of a Workshop on
Extragalactic Radio Sources held in Green Bank, WV, Oct 10-12, 2002,
eds. J.A. Zensus, E. Ros, \& M.H.C. Cohen, in press.





\end{thebibliography}
\end{document}